\documentclass[aps,pra,reprint,superscriptaddress]{revtex4-2}

\usepackage{graphicx}
\usepackage{color}
\usepackage{amsmath,amssymb}
\usepackage{bm}
\usepackage{upgreek}
\usepackage{xspace}

\usepackage{bbold}

\usepackage[english]{babel}
\usepackage[colorlinks,urlcolor=blue,citecolor=blue,linkcolor=blue]{hyperref}
\begin{document}

\title{3D-mapping and manipulation of photocurrent in an optoelectronic diamond device}

\author{A.~A.~Wood}
\email{alexander.wood@unimelb.edu.au}
\affiliation{School of Physics, University of Melbourne, Parkville, Victoria 3010, Australia}
\author{D.~J.~McCloskey}
\affiliation{School of Physics, University of Melbourne, Parkville, Victoria 3010, Australia}
\author{N. Dontschuk}
\affiliation{School of Physics, University of Melbourne, Parkville, Victoria 3010, Australia}
\author{A. Lozovoi}
\affiliation{CUNY-The City College of New York, New York 10031, USA}
\author{R. M. Goldblatt}
\affiliation{School of Physics, University of Melbourne, Parkville, Victoria 3010, Australia}
\author{T. Delord}
\affiliation{CUNY-The City College of New York, New York 10031, USA}
\author{D. A. Broadway}
\affiliation{School of Science, RMIT University, Melbourne Victoria 3000, Australia}
\author{J.-P. Tetienne}
\affiliation{School of Science, RMIT University, Melbourne Victoria 3000, Australia}
\author{B. C. Johnson}
\affiliation{School of Science, RMIT University, Melbourne Victoria 3000, Australia}
\author{K. T. Mitchell}
\affiliation{School of Physics, University of Melbourne, Parkville, Victoria 3010, Australia}
\author{C. T.-K. Lew}
\affiliation{School of Physics, University of Melbourne, Parkville, Victoria 3010, Australia}
\author{C. A. Meriles}
\affiliation{CUNY-The City College of New York, New York 10031, USA}
\affiliation{CUNY - The Graduate Center, New York, NY 10016, USA}
\author{A. M. Martin}
\affiliation{School of Physics, University of Melbourne, Parkville, Victoria 3010, Australia}

\date{\today}
\begin{abstract}
\textbf{Characterising charge transport in a material is central to the understanding of its electrical properties, and can usually only be inferred from bulk measurements of derived quantities such as current flow. Establishing connections between host material impurities and transport properties in emerging electronics materials, such as wide bandgap semiconductors, demands new diagnostic methods tailored to these unique systems, and the presence of optically-active defect centers in these materials offers a non-perturbative, \emph{in-situ} characterisation system. Here, we combine charge-state sensitive optical microscopy and photoelectric detection of nitrogen-vacancy (NV) centres to directly image the flow of charge carriers inside a diamond optoelectronic device, in 3D and with temporal resolution. We optically control the charge state of background impurities inside the diamond on-demand, resulting in drastically different current flow such as filamentary channels nucleating from specific, defective regions of the device. We then optically engineered conducting channels that control carrier flow, key steps towards optically reconfigurable, wide bandgap designer optoelectronics. We anticipate our approach might be extended to probe other wide-bandgap semiconductors (SiC, GaN) relevant to present and emerging electronic technologies.}
\end{abstract}
\maketitle

Characterising a material, however exotic, for use in electronics applications amounts to determining how charges enter, leave, and most importantly, move through it. Imaging charge carriers directly is generally not possible, so an ideal diagnostic platform would thus consist of an array of minimally-perturbative probes embedded throughout a material that detect electrons and holes as they traverse the device. Exactly this situation occurs in several wide bandgap semiconductors, where impurities give rise to optically-active defect centres~\cite{aharonovich_solid-state_2016, atature_material_2018,fedyanin_optoelectronics_2021} distributed throughout the material. Some of these defects even exhibit coherent spin properties at room temperature~\cite{chatterjee_semiconductor_2021}, such as the nitrogen-vacancy (NV) centre in diamond~\cite{doherty_nitrogen-vacancy_2013} or the silicon-vacancy and divacancy in SiC~\cite{koehl_room_2011, kraus_room-temperature_2014}. Recently, it has been shown that exploiting the spin properties of these defects provides a powerful means of characterising optoelectronic devices~\cite{iwasaki_direct_2017, broadway_spatial_2018, forneris_mapping_2018, anderson_electrical_2019, lillie_imaging_2019, zhou_spatiotemporal_2020, zhang_single_2019, scholten_imaging_2022}. However, not all such defects exhibit sufficient sensitivity to electric fields, especially at room temperature. In comparison, the electronic charge state of a deep-level defect in a wide bandgap semiconductor is a stable and optically addressable resource~\cite{jayakumar_optical_2016, dhomkar_long-term_2016} for the study of charge transport~\cite{lozovoi_optical_2021, lozovoi_detection_2023}. In this work, we introduce a new imaging scheme based on single-shot charge-state-sensitive optical microscopy to reveal the generation and flow of currents inside diamond at room temperature in three dimensions. Our technique is distinct from scanning photocurrent microscopy, where optical excitation generates a photocurrent signal at remote electrodes~\cite{ma_photocurrent_2023}: here, we image the actual photocurrent as it flows between electrical contacts and the point of optical illumination, a measurement simultaneously \emph{in-situ} yet also weakly perturbing.  

\begin{figure*}
	\centering
		\includegraphics[width = \textwidth]{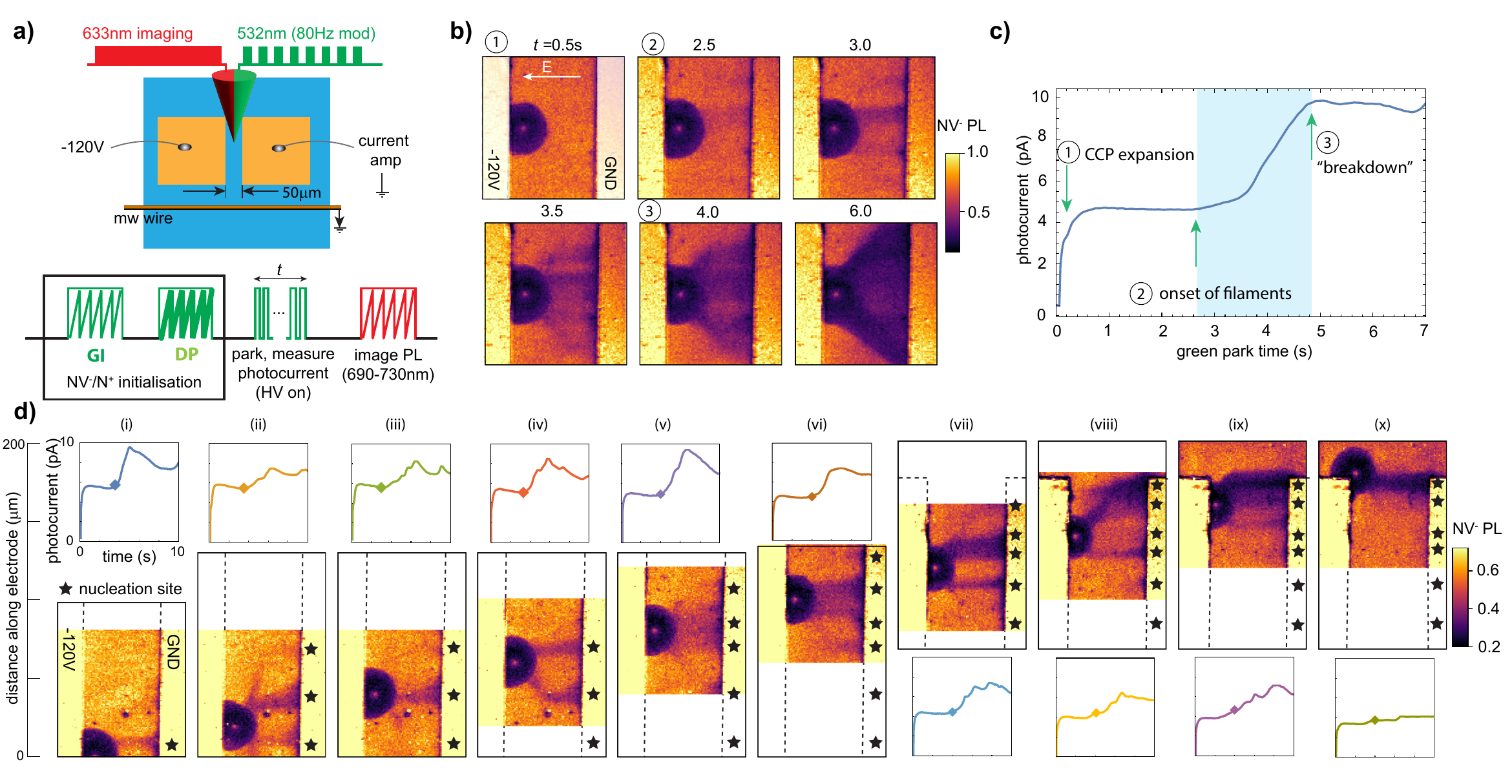}
	\caption{\textbf{Photocurrent generation and detection within an optoelectronic diamond device.} a) Experimental setup: two planar electrodes are deposited on the surface of a diamond hosting NV centres. Optical protocols (bottom frame) are used to prepare the NV centres and other defects in particular charge states, which can then be measured using optical confocal microscopy or detection of photocurrent (PC) via the electrodes and lock-in amplification. (b) Images of NV$^-$ photoluminescence after an 8\,mW (80\,Hz modulated) green laser beam is positioned near the left electrode, held at -120\,V, while the right-hand electrode (positive) connects to ground through a current amplifier, and (c) corresponding photocurrent vs. time. All images are 80$\times$80$\,\upmu$m scans. The formation of dark, NV$^0$-rich filaments connecting the charge capture halo surrounding the illumination spot to the left-hand electrode is correlated with distinct features in the photocurrent, which corresponds to predominantly-hole photocurrent flowing in local channels before a general `breakdown' event at later times. Holes move from right to left in these images. (d)(i-x) The dark `filaments' nucleate from specific regions of the electrodes, as these scans over a 200\,$\upmu$m region shows. Plots of the time-dependent photocurrent are shown inset for each confocal scan, and the coloured circle on each plot indicates the corresponding time where the image is taken (3-4.5\,s, where the filaments are most clearly resolved for each case). At each point, the onset of filaments always precedes a general increase in photocurrent.}
	\label{fig:fig1}
\end{figure*}

Our method works on the basis that a charge carrier induces a measurable change in the fluorescence of a colour centre defect upon capture, for instance NV$^-$ + h$^+ \rightarrow$ NV$^0$. With charge-state selective excitation and spectral filtering, we detect only the NV$^-$ fluorescence, consequently, the spatial profile and time evolution of current flow is imprinted on the charge state of an ensemble of NV centres embedded throughout the device, enabling micoscale mapping in three dimensions. Optical confocal microscopy is augmented by the simultaneous electrical detection of time-dependent currents, allowing us to make clear associations between spatial structures observed in images and current transients. We employ our technique to detect local electric field inhomogeneities and preferential charge injection in a metal-semiconductor interface, revealing points in the metal-diamond junction that nucleate thin filament-like channels of charge carriers. Next, we show that the spatial structure and temporal variation of photocurrent is determined by the charge state of nonfluorescent defects, assumed to be substitutional nitrogen impurities~\cite{ashfold_nitrogen_2020}. We can control and even mimic the naturally-occurring filament structures by creating optically-defined patterns of varied conductivity with laser scans, and then image these patterns in three spatial dimensions with temporal resolution. 

{\bf Experiment.} Our experiment, depicted schematically in Fig. \ref{fig:fig1}(a) and detailed further in Methods, uses a diamond sample equipped with a pair of coplanar Au/Cr electrodes ($500\,\times 500\,\upmu$m, 50$\,\upmu$m spacing) fabricated on the oxygen-terminated diamond surface via photolithography (see Methods). The nitrogen and NV concentrations are $1$\,ppm and $0.01$\,ppm, respectively. A purpose-built confocal microscope enables two-colour optical illumination (532\,nm and 633\,nm) and charge-state-selective imaging via collection of only NV$^-$ fluorescence within a limited band (691-731\,nm). One of the electrodes is held at a potential $V$, while the other connects to ground via a low-noise current preamplifer and lock-in amplifier in a standard setup to detect photocurrent~\cite{siyushev_photoelectrical_2019, bourgeois_photoelectric_2020}. The voltage-current characteristics of the device were measured and found to exhibit the characteristic back-to-back Schottky-diode shape of similar semiconductor-metal interface devices~\cite{grillo_currentvoltage_2021}, preventing efficient injection of charge into the diamond from the electrodes. We thus generated a photocurrent inside the diamond using optical illumination and charge cycling~\cite{aslam_photo-induced_2013} of the NV centres, which injects free electrons and holes into the conduction and valence bands of the diamond, respectively. These freely-diffusing charge carriers are then trapped by various defects, or drift towards the electrodes and detected as a photocurrent~\cite{bourgeois_photoelectric_2015}. In the former case, and in the absence of an electric field, positively-charged holes are efficiently captured by negatively-charged NV$^-$ centres, resulting in a characteristic `halo'-like charge capture pattern (CCP) of NV$^0$ and thus reduced fluorescence around the illumination point in an NV$^-$-selective image. Combining photoelectric detection with charge-sensitive imaging allows us to then track where photocurrent flows when an electric field is applied. 

Prior to inducing the photocurrent, we utilise two protocols, \emph{green initialisation} (GI) and \emph{defocused preparation} (DP), depicted in Fig. \ref{fig:fig1}(a), to initialise the charge state of both the NV centres and nitrogen in the diamond into NV$^-$/N$^+$ (see Methods). While other defects are expected to play a similar role~\cite{zheng_coherence_2022}, we focus on nitrogen given it is the highest abundance impurity in our sample. A full understanding of the charge environment we initialise necessarily includes consideration of other defects, including vacancies, as discussed later. After initialisation, we apply a -120\,V bias voltage to the left-hand electrode and position a focused green laser beam (8\,mW, $1\,\upmu$m diameter) near the same electrode for a variable duration. The position of the laser beam was chosen to maximise the measured steady-state photocurrent (see Supplementary Material). The green light is modulated at 80\,Hz to facilitate lock-in detection of the photocurrent, the modulation itself was not observed to have any confounding effects on the measurements. While the measured photocurrent originates from various sources~\cite{hruby_magnetic_2022}, we verified that a significant fraction derives from NV charge cycling by performing a photoelectrically-detected magnetic resonance measurement with an applied microwave field (see Supplementary material). 

In Fig. \ref{fig:fig1}(b,c), we study the time dependence of the NV population and photocurrent, respectively. We first observe the characteristic circular CCP of NV$^0$ that quickly grows to a radius of 10-15$\,\upmu$m within the first 0.5\,s of the photocurrent laser application. This feature expands almost isotropically regardless of the applied field and is driven by charge carrier diffusion and space charge effects~\cite{lozovoi_probing_2020}. At later times, we observe the formation of filamentary structures of NV$^0$ that connect the CCP to the ground (positive) electrode. The formation of these features correlates with sudden increases in photocurrent, indicating that hole photocurrent flows from the ground electrode into the CCP via these channel-like filaments. Later, a significant `breakdown'-like event is observed, corresponding to an almost doubled photocurrent and complicated, unstable temporal dynamics. We also studied how the photocurrent and filaments depend on position of the laser, laser power and applied electric field (see Supplementary material). 

{\bf In-situ device characterisation.} The filamentary current channels are observed to nucleate from specific regions of the ground (positive) electrode, and grow from right to left, starting as dark patches near the electrode that eventually reach out to join the CCP. In some cases, a matching filament from the CCP is observed to join the approaching filament from the electrode. In Fig. \ref{fig:fig1}(d) we shift the vertical position where the photocurrent beam is applied across the full 200\,$\upmu$m field of view of our microscope, and observe consistent sites where the filaments form, sometimes taking nontrivial paths [e.g. Fig. \ref{fig:fig1}(d, ii, iv)] to join to the source CCP. The onset of filaments is always accompanied by a steep increase in photocurrent (inset). 

A hint to the origin of these photocurrent filament channels is found in the observed nucleation points along the electrode, and a consistent nucleation point appears to be the top corner of the electrode, where field bunching results in a locally stronger electric field. We hypothesise then that filaments nucleate at similar points of higher field along the electrode, whether due to localised intrinsic interface defects and damage or imperfections in the metal-semiconductor junction. This latter point is of considerable interest, given the difficulty in fabricating electrical contacts on diamond and the lack of existing characterisation techniques. We note that for all applied electric fields in this work, estimated to be of order $10$-$20\,$kV\,cm$^{-1}$, NV electrometry~\cite{dolde_electric-field_2011, michl_robust_2019, mittiga_imaging_2018} could not detect any electric-field induced splittings or shifts. In contrast, optical measurement of the photocurrent via NV charge state conversion yields a striking and highly specific \emph{in-situ} detection method for local electric field inhomogeneity.

\begin{figure}
	\centering
		\includegraphics[width = \columnwidth]{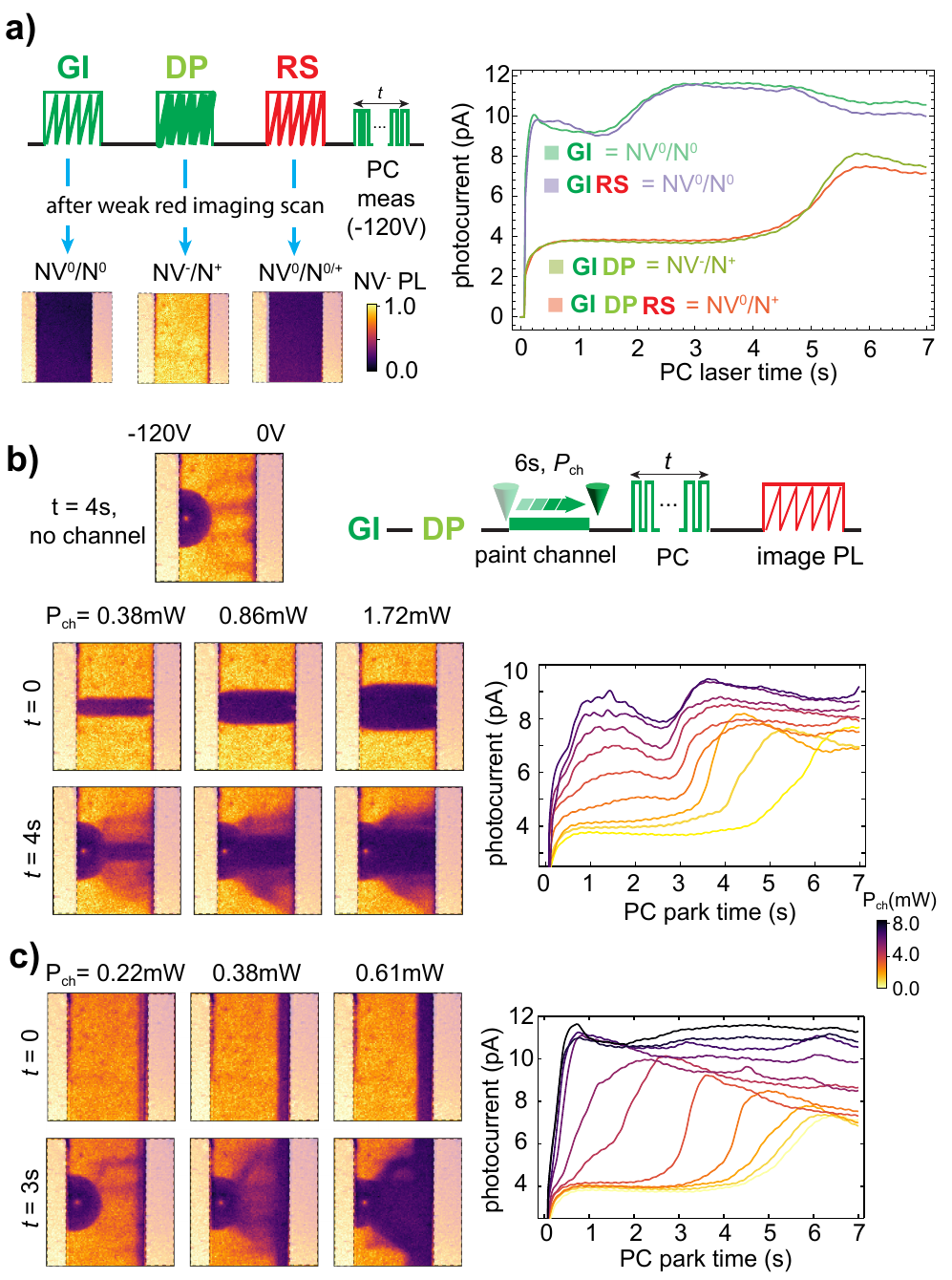}
	\caption{\textbf{Optically engineering the initial conductivity.} (a) Three optical preparation steps control the charge state of NV centres and nitrogen impurities between the electrodes. GI prepares mostly NV$^0$ and N$^0$ due to charge cycling and carrier capture, DP creates NV$^-$ and N$^+$ while an RS scan ionises NV$^-$ and leaves N$^0$ and N$^+$ untouched. Photocurrent measurements following initialisation steps are independent of the NV charge state, but strongly dependent on the assumed N charge state. (b) Drawing horizontal or (c) vertical paths with a varied-power green laser defines conductive channels, with wider horizontal channels resulting in greater photocurrent generation, and wider vertical channels parallel to the negative electrode nucleating filaments at earlier times, with concomitantly higher photocurrents.}
	\label{fig:fig2}
\end{figure}

{\bf Tailoring photoconductivity with dark impurities.} We now draw special attention to the role of charge-state preparation, not just of the NV centres but of other, non-luminescent defects in the diamond. Indeed, for the charge state of the NV centres to be a non-perturbative detector carrier transport, the extent to which the probes affect the photocurrent dynamics must first be understood. We first remove the DP step, which generates the NV$^-$ charge state, resulting in initialisation into optically dark NV$^0$/N$^0$ (see also Methods). Figure \ref{fig:fig2}(a) shows the photocurrent as a function of time following initialisation with and without DP, with almost twice as much photocurrent observed without the DP step, and lacking the intervening filament nucleation step. While this might initially point to the NV charge state playing a significant role in photocurrent transport, further investigation reveals the opposite. We introduce another step, consisting of a 300\,$\upmu$W red (633\,nm) scan (RS) that follows either GI or DP. Illumination with weak red light results in one-way ionisation of the NV$^-$ to NV$^0$~\cite{wood_wavelength_2024}, but with a photon energy of 1.96\,eV, below the 2.2\,eV photoionisation threshold of N$^0$~\cite{bourgeois_enhanced_2017}, the red light leaves the nitrogen charge state unchanged.

In Fig. \ref{fig:fig2}(a) we compare the previous intialisation stages with the addition of a RS step. Notably, the photocurrent is unchanged following initialisation to NV$^-$ and subsequent ionisation with red scanning, and essentially identical when a red scan follows a GI-only initialisation. The nearly identical photocurrent profiles correlate with the expected nitrogen charge state and confirm the NV plays a negligible role in the photocurrent dynamics, serving only to detect the spatial distribution of holes.   

Photocurrent appears to flow readily in N$^0$-rich regions, and less so in N$^+$-rich regions. Rather than scanning across the whole inter-electrode region as in a GI step, we can optically engineer specific regions by `painting' patterns of NV$^0$/N$^0$ with green light. In Fig. \ref{fig:fig2}(b), we scan a green laser beam with power $P_\text{ch}$ between the electrodes, making a horizontal conducting channel with width set by $P_\text{ch}$: as the channel is widened with increasing power, the time taken for the photocurrent to reach the same saturation value decreases. The time-dependence mirrors the signatures of filament formation, with a characteristic step in photocurrent signal occurring at earlier times as the channel width, effectively an artificial filament, is increased. Similarly, in Fig. \ref{fig:fig2}(c), we paint a vertical line parallel to the right-hand electrode. Now, filaments form between the source CCP and the painted region, accompanied by the same photocurrent signatures observed for filaments formed between the CCP and the bare electrode. Interestingly, this observation implies the processes that drive filament nucleation at the diamond-electrode interface are replicated in the defect charge environment following green laser application, which we assume generates N$^0$ in its vicinity.   

\begin{figure*}
	\centering
		\includegraphics[width = \textwidth]{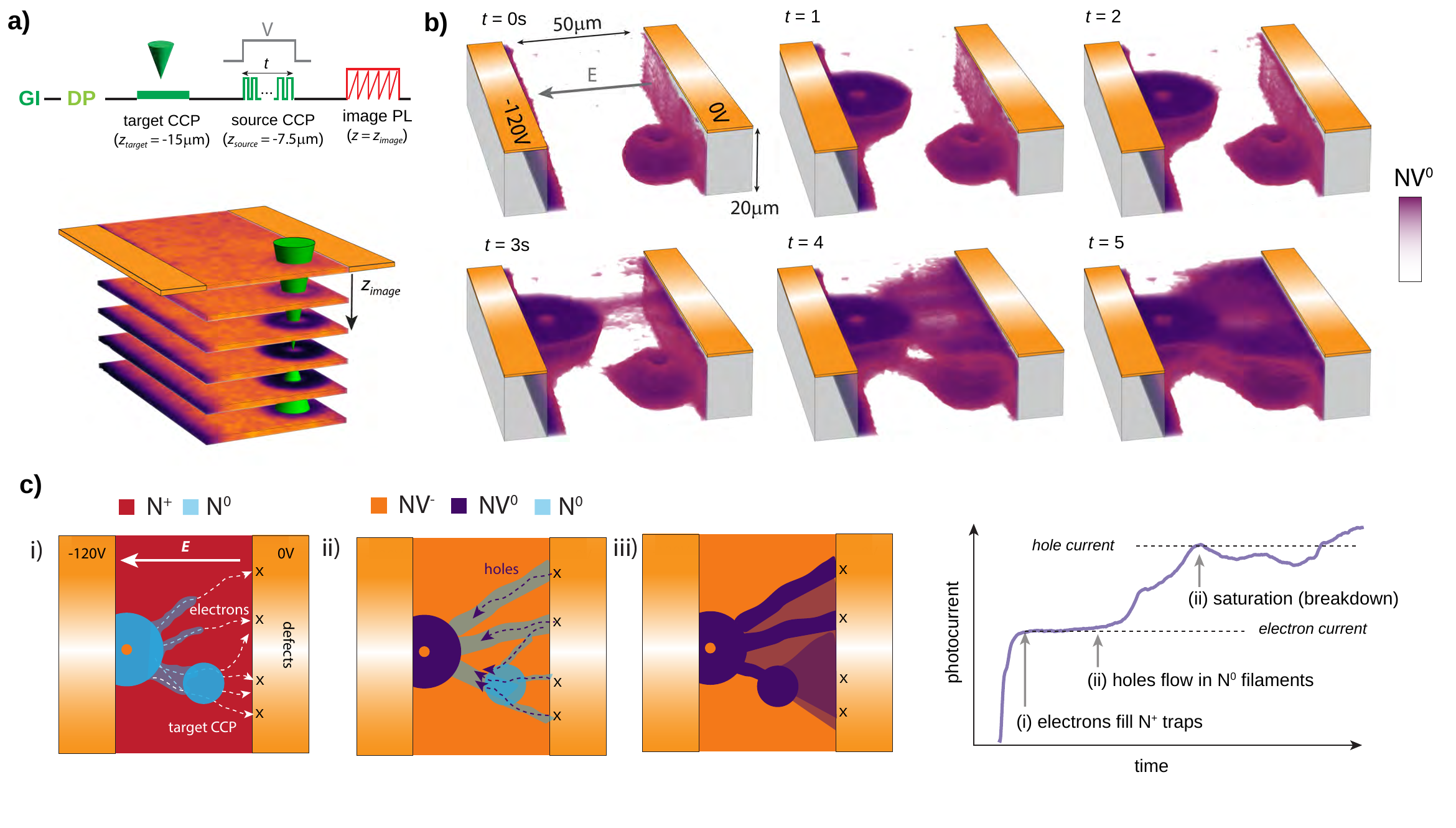}
	\caption{\textbf{3D-spatiotemporal mapping of photocurrent flow.} (a) Illustration (bottom) and protocol (top) for time-resolved 3D photocurrent imaging with charge sensitive microscopy. Here, a `target' CCP is generated with a 0.50\,s, 1.5\,mW laser park 15\,$\upmu$m below the diamond surface. Varying the $z$-plane where the $x$-$y$ confocal scan is performed reveals the near-spherical shape of the target CCP, which is fully disconnected from the metal electrodes (the dark sections on the sides are the result of shadows cast by the electrodes). (b) Applying an opacity mask to display only the NV$^0$ concentration, we reveal the full 3D spatiotemporal dynamics of photocurrent generation, filamentation and breakdown. The source photocurrent laser is applied for time $t$, 7.5\,$\upmu$m below the surface, the uppermost $z$-position corresponds to $z = 5\,\upmu$m below the diamond surface. We note the distortion and eventual filament nucleation from the target CCP, 15\,$\upmu$m below the surface. (c) A simple cartoon model of filament formation and corresponding photocurrent accumulation. i), electrons generated by the photocurrent laser drift under the applied field, being captured by N$^+$ and consequently forming channels of N$^0$ (ii) that act as conduction pathways for holes, at later times individual filaments are washed out and hole flow saturates (iii). The proposed role of the target CCP in this picture is to gather and bunch hole trajectories towards the CCP, which explains the bulging observed in (b).}
	\label{fig:fig3}
\end{figure*}    

{\bf 3D imaging of photocurrent flow.} 
The electrical conductivity of diamond is strongly dependent on the surface chemistry~\cite{pakes_diamond_2014}, which can also be modified by light~\cite{zheng_coherence_2022}. An important question then arises as to whether the surface of the diamond plays a significant role in our own experiments. Due to the lateral electrode geometry of our device, the electric field strength decreases linearly with depth (see Supplementary Material), meaning the strongest carrier drift is found at or near the surface. We extended our photocurrent imaging scheme into three dimensions (Fig. \ref{fig:fig3}(a)) by incrementally changing the $z$-plane where the red readout scan is performed, with all other optical manipulation steps as before. To probe the depth dependence of photocurrent transport, we first apply a green laser beam ($P = 1.2$\,mW, for 0.5\,s) 15\,$\upmu$m below the surface of the diamond and $10\,\upmu$m from the ground electrode to generate an isolated, `target' CCP, analogous to the painted regions depicted in Fig. \ref{fig:fig2}(b,c) though entirely separate from the electrodes. 3D charge-sensitive confocal scanning (Fig. \ref{fig:fig3}(a,b)) reveals the characteristic shape of the CCP, consisting of a spherical region of dark NV$^0$ punctured by the conical regions exposed to the entering and exiting laser beam, which remain NV$^-$~\cite{wood_wavelength_2024}. 

With the target CCP in place, we made a photocurrent measurement as before, with the `source' laser positioned 7.5$\upmu$m below the surface for a varied time $t_p$. Figure \ref{fig:fig3}(b) shows a sequence of 3D maps of the resulting NV$^0$ distributions following variable time photocurrent laser measurements. As time progresses, we see filament formation extending more than 10\,$\upmu$m into the diamond bulk prior to the onset of photocurrent breakdown. More significantly, we observe a distinct, isolated filament form to join the source and target CCP. These measurements show unambiguously that the photocurrent spatial dynamics result from charged defects inside the diamond bulk, rather than surface conduction effects. Prior to formation of a connecting filament, the target CCP bulges \emph{toward} the photocurrent-induced CCP (see accompanying videos in the Extended Data). This observation, while in concordance to some degree with our measurements showing filament nucleation from both the bare electrodes and the `painted' regions in Fig. \ref{fig:fig2}(c), is unexplained, as the `target' CCP is assumed to be fully disconnected from the ground electrode.  

{\bf Discussion.} Our measurements provide an intricate picture of the origin and phenomenology of photocurrent propagation in diamond. We first note that holes appear to cross the metal-diamond junction of the ground electrode and converge towards the CCP, first as channel-like filaments and then as a breakdown-like flow that converges towards the CCP in a broad ``V"-like shape. To understand this, we must again take into consideration that the charge state of nonfluorescent defects has a significant effect on the spatio-temporal dynamics of the current flow. While holes travel from right to left in images such as Fig. \ref{fig:fig1}(b) and leave visible evidence in the form of NV$^0$, electrons are to be assumed to follow similar, unseen trajectories but from left to right. Electrons generated by the photocurrent laser remain invisible to our imaging technique, as the dominant interaction expected is capture by N$^+$, which does not change the NV fluorescence, but does significantly alter the photoconductivity. From Fig. \ref{fig:fig1}(d), hole injection is observed to occur at specific, reproducible regions of the ground electrode, and we expect that these same sites also attract electron flow towards regions of stronger electric field. Electrons create channels of N$^0$ as they are drawn to these defective sites, and then build up on the diamond-electrode Schottky interface. Eventually, sufficient negative charge buildup occurs to reduce the Schottky barrier height and facilitate hole (minority carrier) injection~\cite{green_minority_1973}, which in turn preferentially flows along the channels passivated by the electrons.

We can thus posit a model where charged impurity scattering sets the electron and hole mobilities, and the filaments are first traced by electrons propagating away from the photocurrent laser beam, as depicted in Fig. \ref{fig:fig3}(c). Open questions remain as to the microscopic charge environment, and whether the nitrogen is initialised into a positive charge state alone (as an optically-defined space charge region), or compensated by electron capture by neutral defects such as neutral vacancies V$^0$~\cite{davies_vacancy_1993}. In the latter case, the charge environment is neutral, though possesses equal numbers of ionised impurities acting as electron and hole traps, resulting in the observed impeded current flow. In both pictures, carrier capture drives the charge environment towards a state composed of neutrally-charged defects (N$^0$, V$^0$) which results in higher carrier mobilities. The time-dependence of the photocurrent, which resembles the archetypal case for extrinsic, defect-mediated photoconductivity~\cite{boer_semiconductor_2018} offers further insights. Given a significant photocurrent flows before any filaments are observed optically, we can deduce that the initial photocurrent is due to electrons, which gradually passivate the N$^+$ population that impedes hole flow. With a high hole mobility corridor of N$^0$ connecting the CCP to the electrode, hole current now flows and is observed through charge state conversion of the NV centres.  

We now apply our model to explain the phenomenology observed in Fig. \ref{fig:fig2}(c) and Fig. \ref{fig:fig3}(b). Of particular note is the somewhat anomalous finding that an isolated target CCP appears to bulge or distort in the direction of the photocurrent CCP. We note that no bulging or distortion is observed for an equivalent experiment without the applied electric field, or without the photocurrent laser but with an applied field. Due to the the very short carrier lifetime ($\sim 1\,\upmu$s~\cite{grivickas_carrier_2020}), no free charge is stored in the isolated target CCP, which is created several seconds before the photocurrent beam is turned on. We propose that the isolated target CCP acts essentially as a focal point first for electron current, which is drawn to the isolated region of NV$^0$/N$^0$ in the same way as it is focused on defective sites along the electrode and the painted regions in Fig. \ref{fig:fig2}(c). This in turn attracts holes, which flow along various defect-nucleated paths from the electrode towards the target CCP. Here, the paths bunch up, resulting in a higher hole concentration and stronger NV$^0$ signal than closer to the positive electrode. Further work is needed to fully understand this observation.   

Our work demonstrates a powerful new measurement modality for the study of charge transport in a semiconductor, and poses questions that motivate an improved understanding of photoelectric physics in diamond. In this work we have focused on diamond, which is arguably the most technologically mature platform for our technique, possessing well-characterised photochromic defects which are stable at room-temperature, while also being a promising material for select high-power electronics~\cite{wort_diamond_2008, donato_diamond_2019, araujo_diamond_2021}. With further work, our method could be extended to materials more advanced in terms of power-efficient semiconductor applications which host similar optically active defect centres~\cite{zhang_material_2020}, i.e., SiC~\cite{niethammer_coherent_2019} or GaN~\cite{zhou_room_2018}, or in emerging platforms for quantum sensing such as hBN~\cite{tran_quantum_2016}. Focusing on diamond, our work has immediate implications for photoelectric detection of NV centres~\cite{zheng_electrical-readout_2022, morishita_spin-dependent_2023}. Photoelectric detection of NV magnetic resonance promises fully-integrated sensors unconstrained by the diffraction limit and limitations of optical collection efficiency~\cite{zheng_electrical-readout_2022}. Indeed, the complicated spatial dynamics of photocurrent we measure here are at play, unseen, in many photoelectric detection schemes. Similar issues are encountered in diamond radiation detectors, where impurities capture charge carriers generated by high-energy particles~\cite{shimaoka_recent_2022}, leading to space-charge effects that impact performance. However, photocurrent cannot be spectrally filtered like photoluminescence, and the carrier capture by other defects we observe here translates to reduced photocurrent yield. Both these points are addressed in our work, where we show careful control of the nitrogen charge state yields significantly enhanced photocurrents, and observation of filament formation allows us to attribute sources of photocurrent injection to from the electrodes rather than from particular defects. More ambitiously, optical manipulation of specific defect concentrations and regions could be used to create junctions, current pathways, interconnects and potentially even entire optically-defined devices using charged defects. 

\section{Methods}
\subsection{Initialisation scans.}
{\bf Green initialisation (GI)} An initial high-intensity (8\,mw) green scan serves to erase any spatial charge distributions left by preceding experiments, and due to the continuous generation of electrons and holes from NV charge cycling originating from the scanning laser beam, prepares a region of predominantly NV$^0$ (via NV$^-$ hole capture) and N$^0$ (via N$^+$ electron capture)~\cite{dhomkar_-demand_2018}. Typical scan ranges are $(80\times80)\,\upmu$m and take 30\,s to complete. The electric field is off during the initialisation scans.

{\bf Defocused preparation (DP)} Following the (GI) scan, we elevate the microscope objective 50\,$\upmu$m above the surface of the diamond and repeat the scan. The corresponding optical intensity at the diamond is significantly reduced, and as a result two-photon NV charge cycling is markedly suppressed. However, single-photon ionisation of N$^0$ continues. Following this defocused preparation (DP) step, we observe an increased fraction of NV$^-$, which we attribute to photoionisation of N$^0$ generating electrons that are eventually captured by the NV$^0$ defects. The extended optical interaction time ($>10\,$s) prevents electron recapture by N$^+$ and effectively forces electrons into traps, resulting in a higher fraction of positively charged nitrogen. More details about how the DP step functions to increase N$^+$ density are provided in the Supplementary Information.

\subsection{Electrode fabrication.}
Square $500\,\times500\,\upmu$m electrodes were fabricated following a photolithographic patterning and lift-off process. The diamond substrate was first cleaned by boiling in a 1:20 (by weight) mixture of sodium nitrate and sulphuric acid for 20 minutes followed by a 10 minute immersion in piranha etching solution at 90 degrees C and subsequent ultrasonic immersion in deionized water. The sample was coated with image-reversal photoresist (AZ1514E – MicroChemicals GmbH) by spin-coating. After baking to remove resist solvent, the sample underwent a brief flood exposure under a standard mercury vapor lamp in a mask aligner (Quintel), after which the sample underwent image-reversal baking in order to partially crosslink the uppermost layer of the resist film. Electrode patterns were then written on the sample using a direct-write UV lithography system (POLOS NanoWriter), after which the resist was developed in AZ 726 MIF developer for 90 seconds. The crosslinked upper layer of resist results in an undercut sidewall profile, thereby allowing for evaporation and lift-off to create the electrodes. The sample was exposed to oxygen-argon plasma in a benchtop plasma ashing system (Diener Femto) for 30 seconds to remove photoresist residues from the sample surface prior to electron-beam evaporation of a 15nm/100nm Cr/Au stack. Lift-off was performed by ultrasonication in acetone followed by deionized water. No post-fabrication annealing of the electrodes was performed, and the sample preparation steps leave the diamond with an oxygen-terminated surface.

\section*{Acknowledgments}
We thank R. E. Scholten for a careful reading of the manuscript. This work was supported by the Australian Research Council (DE210101093). R. M. G. and K. T. M. were supported by an Australian Governement Research Training Program (RTP) Scholarship. D. A. B. was supported by an ARC DECRA Fellowship (DE230100192). A.L. acknowledges support from the National Science Foundation under grant NSF-2216838. T.D. and C.A.M. acknowledge support by the U.S. Department of Energy, Office of Science, National Quantum Information Science Research Centers, Co-design Center for Quantum Advantage (C2QA) under contract number DE-SC0012704. A.L., T.D., and C.A.M. acknowledge access to the facilities and research infrastructure of NSF CREST-IDEALS, grant number NSFHRD- 1547830. 

\section*{Author contributions}
A. A. W. conceived and led the work with guidance from C. A. M. and A. M. M. Data was collected and analysed by A. A. W. with assistance by R. M. G. The devices were fabricated by D. J. M., simulations of the device electric fields were carried out by  K. T. M, and the electrical characteristics of the device were assisted by C. T-K. L. The theoretical model was developed by A. A. W, N. D., D. J. M., A. L., T. D. and C. A. M. with inputs from D. A. B., J-P. T. and B. C. J. All authors contributed to the preparation of the manuscript. 

\section*{Competing interests}
The authors declare no competing interests.

\section*{Data availability}
All raw data and analysis code is available from the corresponding author (A. A. W.) upon reasonable request.


%%%%%%%%%% Merge with supplemental materials %%%%%%%%%%
\clearpage
\widetext
\begin{center}
\textbf{\Large SUPPLEMENTARY INFORMATION}
\end{center}
%%%%%%%%%% Merge with supplemental materials %%%%%%%%%%
%%%%%%%%%% Prefix a "S" to all equations, figures, tables and reset the counter %%%%%%%%%%
\setcounter{equation}{0}
\setcounter{figure}{0}
\setcounter{table}{0}
\setcounter{page}{1}
\makeatletter
\renewcommand{\theequation}{S\arabic{equation}}
\renewcommand{\thefigure}{S\arabic{figure}}
\renewcommand{\bibnumfmt}[1]{[S#1]}
\renewcommand{\citenumfont}[1]{S#1}
%%%%%%%%%% Prefix a "S" to all equations, figures, tables and reset the counter %%%%%%%%%%

\section{Experimental Setup}
\begin{figure}[t!]
	\centering
		\includegraphics[width = \textwidth]{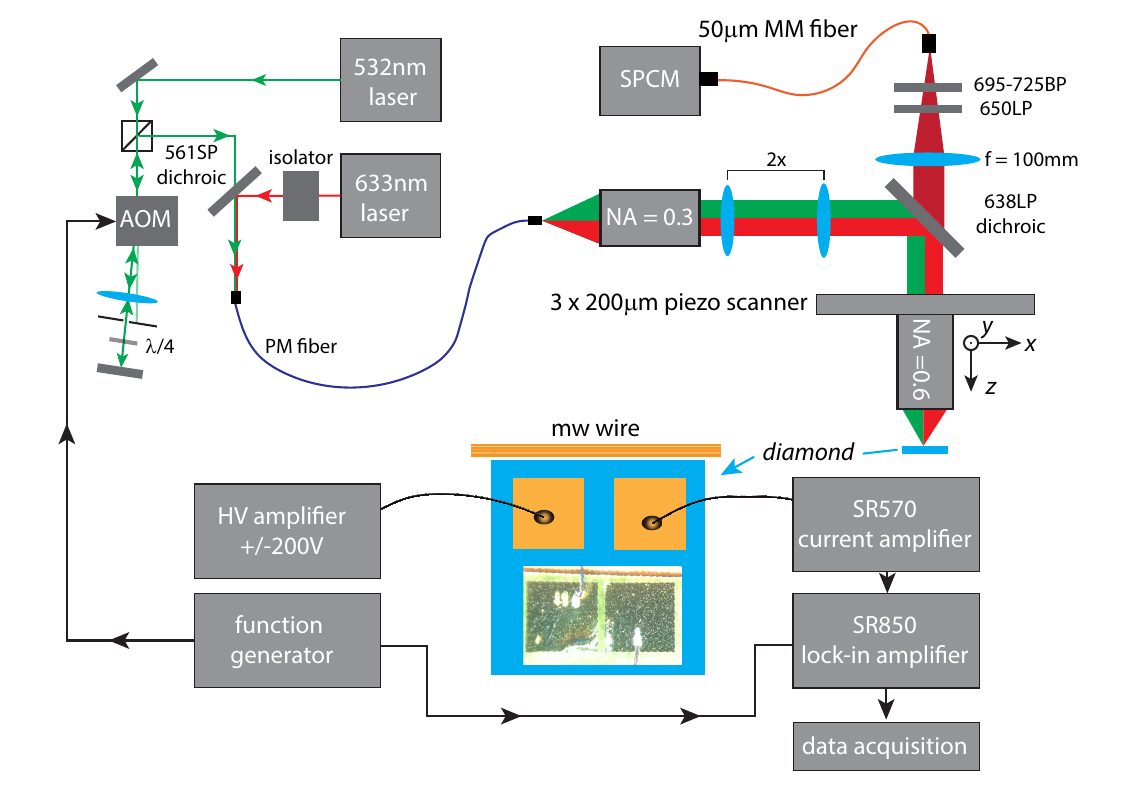}
	\caption{Experimental setup, showing relevant experimental hardware and geometry. Inset on the schematic diagram of the diamond is a photograph of the actual device, the electrode pads are $500\,\times500\,\upmu$m for scale.}
	\label{fig:figs1}
\end{figure}

The experimental setup was briefly outlined in the main text and methods, and a full description including all important part numbers is provided here and shown schematically in Fig. \ref{fig:figs1}. The apparatus is a home-built confocal microscope featuring a three-axis scanning microscope objective and paths for multiple wavelength excitation, and a single-photon counting module for detection of NV fluorescence. We use a Nikon TU Plan ELWD 50$\times$ NA = $0.6$ microscope objective and a Physik Instrumente P-545 scanning piezo stage with 200$\,\upmu$m travel in each direction. Excitation light at 633\,nm from a diode laser (Coherent Obis 633) and 532\,nm from a Laser Quantum opus 532 via a double-passed AOM setup (AA Optoelectronic MT200-A0,5-VIS) for intensity control is combined with free space optics and dichroic mirrors into a single-mode, polarisation-maintaining fiber (Thorlabs P3-488PM-FC). Light from the fiber is collimated with an achromatic microscope objective (Nikon TU Plan EPI SLWD 0.3 NA) and expanded to fill the back aperture of the imaging objective (5\,mm). A Thorlabs 638\,nm longpass dichroic mirror reflects green and red light onto the microscope objective and permits the transmission of NV fluorescence. A 650\,nm longpass filter (Thorlabs) and a 695-725\,nm bandpass filter (Semrock) set of optical filters is used to selectively transmit only NV$^-$ fluorescence via an $f$ = 100\,mm doublet lens into 50\,$\upmu$m multi-mode fiber, and inally into a single-photon counting module (SPCM, Excelitas SPCM-AQRH-14). Despite the filters, we still detect residual 633\,nm light from surface reflections in our signal, though at the level of a few thousand counts per second can be generally ignored, contributing only a baseline background signal to the NV$^0$ signal. 

The diamond sample is mounted using PDMS on a microscope cover slip glued onto a custom printed circuit board. Care is taken to avoid any glue (Norland Optical Adhesive no. 61) contacting electrical pads as it has been found to exhibit conductivity. Square $500\,\times500\,\upmu$m electrodes were fabricated following a photolithographic patterning and lift-off process. The diamond substrate was first cleaned by boiling in a 1:20 (by weight) mixture of sodium nitrate and sulphuric acid for 20 minutes followed by a 10 minute immersion in piranha etching solution at 90 degrees C and subsequent ultrasonic immersion in deionized water. The sample was coated with image-reversal photoresist (AZ1514E – MicroChemicals GmbH) by spin-coating. After baking to remove resist solvent, the sample underwent a brief flood exposure under a standard mercury vapor lamp in a mask aligner (Quintel), after which the sample underwent image-reversal baking in order to partially crosslink the uppermost layer of the resist film. Electrode patterns were then written on the sample using a direct-write UV lithography system (POLOS NanoWriter), after which the resist was developed in AZ 726 MIF developer for 90 seconds. The crosslinked upper layer of resist resulted in an undercut sidewall profile, thereby allowing for evaporation and lift-off to create the electrodes. The sample was exposed to oxygen-argon plasma in a benchtop plasma ashing system (Diener Femto) for 30 seconds to remove photoresist residues from the sample surface prior to electron-beam evaporation of a 15nm/100nm Cr/Au stack. Lift-off was performed by ultrasonication in acetone followed by deionized water. No post-fabrication annealing of the electrodes was performed, and the sample preparation steps leave the diamond with an oxygen-terminated surface.

The diamond electrodes are connected to the PCB via wirebonds, and an electrically-insulated 100\,$\upmu$m copper wire is placed 100\,$\upmu$m above the electrodes for the application of microwaves, which are sourced from a Windfreak SynthHD connected to a Microwave Amplifiers AM6-1-3.5-47-47 50\,W amplifier. We use a high-voltage amplifier (Thorlabs HVA200) to supply voltages to one electrode between $\pm200\,$V during usual operation or a high-voltage DC source (Stanford PS325) for higher voltages (up to $\pm500\,$V) for I-V characterisation. High-voltage control signals are sourced from a 16-bit analog voltage source (NI-PCI6733). The second electrode is connected to the photocurrent signal measurement chain, consisting of a low-noise current preamplifier (Stanford SR570) whose output is connected to a lock-in amplifier (Stanford SR850), the output of which is directed to a National Instruments data acquisition board (NI-PCIe-6323). For photocurrent measurements, the lock-in reference signal is provided by a Rigol DG4162 function generator that modulates the AOM drive amplitude in the 532\,nm double-pass setup. Typical gain settings for the current amplifier are 100-500\,pA/V, no filtering, high-BW mode and no input offsets. The lock-in amplifier has 20\,dB of gain (the lowest setting) and a 30\,ms time constant, with the modulation frequency set to 80\,Hz, square-wave. We do not require the lock-in amplifier to see a photocurrent signal, but find it essential to remove anomalous drifts in the observed DC signal following optical preparation steps, where a slowly-damped (60\,s) DC component of between 10-100\,pA often accompanies the photocurrent signal, discussed further in Sec. \ref{sec:dev_chars}. The origin of this effect is not precisely known, though is potentially related to the unavoidable optical excitation of the metal-diamond interface during initialisation scans\cite{rieger_fast_2023s}.  Our experimental hardware and data analysis is controlled using the Python-based Qudi experimental control suite~\cite{binder_qudi_2017s}. 

\section{Photoelectrically-detected magnetic resonance}
\begin{figure}[t!]
	\centering
		\includegraphics[width = \textwidth]{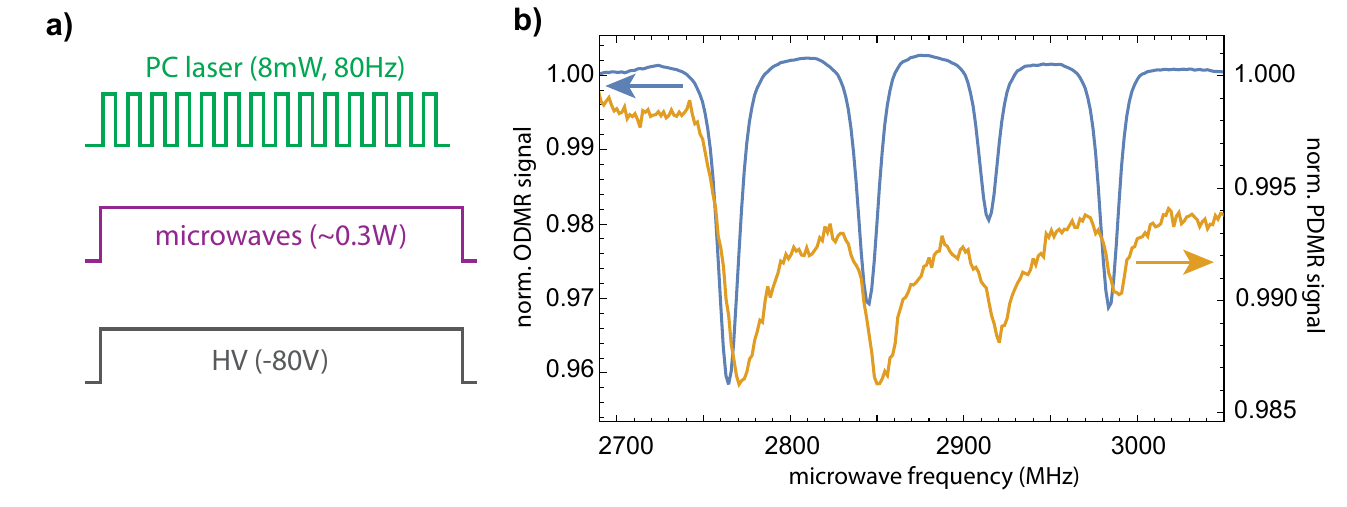}
	\caption{Experimental protocol (a) for photoelectrically-detected magnetic resonance (PDMR), consisting of 300\,mW varied-frequency cw microwave and a -80\,V electric field applied while the 8\,mW 532\,nm laser is modulated at 80\,Hz. (b) ODMR (blue, left-axis) and PDMR (orange, right axis) results. The slight offset between ODMR and PDMR traces is due to the long time constant of the lock in amplifier making the photocurrent lag behind the time-varying microwave frequency sweep. }
	\label{fig:figs2}
\end{figure}

We confirmed that the measured photocurrent originates from charge cycling of the NV by performing photoelectrically-detected magnetic resonance (PDMR). In a PDMR experiment, a change in photocurrent signal is observed when microwaves resonantly drive the $|m_S = 0\rangle\leftrightarrow |m_S = \pm 1\rangle$ transitions. This originates from the spin-selectivity of the NV$^-$ ionisation process, as green-mediated ionisation is more probable from the excited state spin triplet of the NV rather than the metastable spin singlet states that are populated preferentially by $m_S = \pm1$ states following optical excitation and passage through the intersystem crossing~\cite{bourgeois_photoelectric_2020s}. A similar effect has also been observed for charge-capture patterns (CCPs), since equivalent physics drives the expansion of the NV$^0$ region around the illumination spot~\cite{jayakumar_long-term_2020s}. Here, observation of PDMR serves primarily to confirm that NV centres are the source for a fraction of the photocurrent. To observe PDMR, a 50\,G magnetic field is applied that makes a 21.1$^\circ$ angle to the diamond surface normal, and we apply the protocol depicted in Fig. \ref{fig:figs2}(a), identical to a cw-ODMR experiment except that the 8\,mW green laser is modulated at 80\,Hz to enable lock-in detection of the photocurrent. PDMR (orange) and ODMR (blue) traces are shown in Fig. \ref{fig:figs2}(b). The observed PDMR contrast is typically less than one percent, and sits on a strong background variation that persists when the laser is switched off, we attribute this to coupling of the microwave signal into the photocurrent detection signal chain. The slight offset between ODMR and PDMR traces is due to the long time constant of the lock in amplifier making the photocurrent lag behind the time-varying microwave frequency sweep. The observation of PDMR indicates a fraction of the photocurrent originates from charge cycling of NV centres. While negative ODMR contrast is the norm, PDMR contrast can vary significantly in amplitude and even become positive for certain samples~\cite{bourgeois_photoelectric_2022s}. Thus, the low amplitude of the PDMR signal does not necessarily indicate a very small photocurrent of purely NV origin.  

\section{Electric field geometry and device characterisation}\label{sec:dev_chars}

\begin{figure}[t!]
	\centering
		\includegraphics[width = \textwidth]{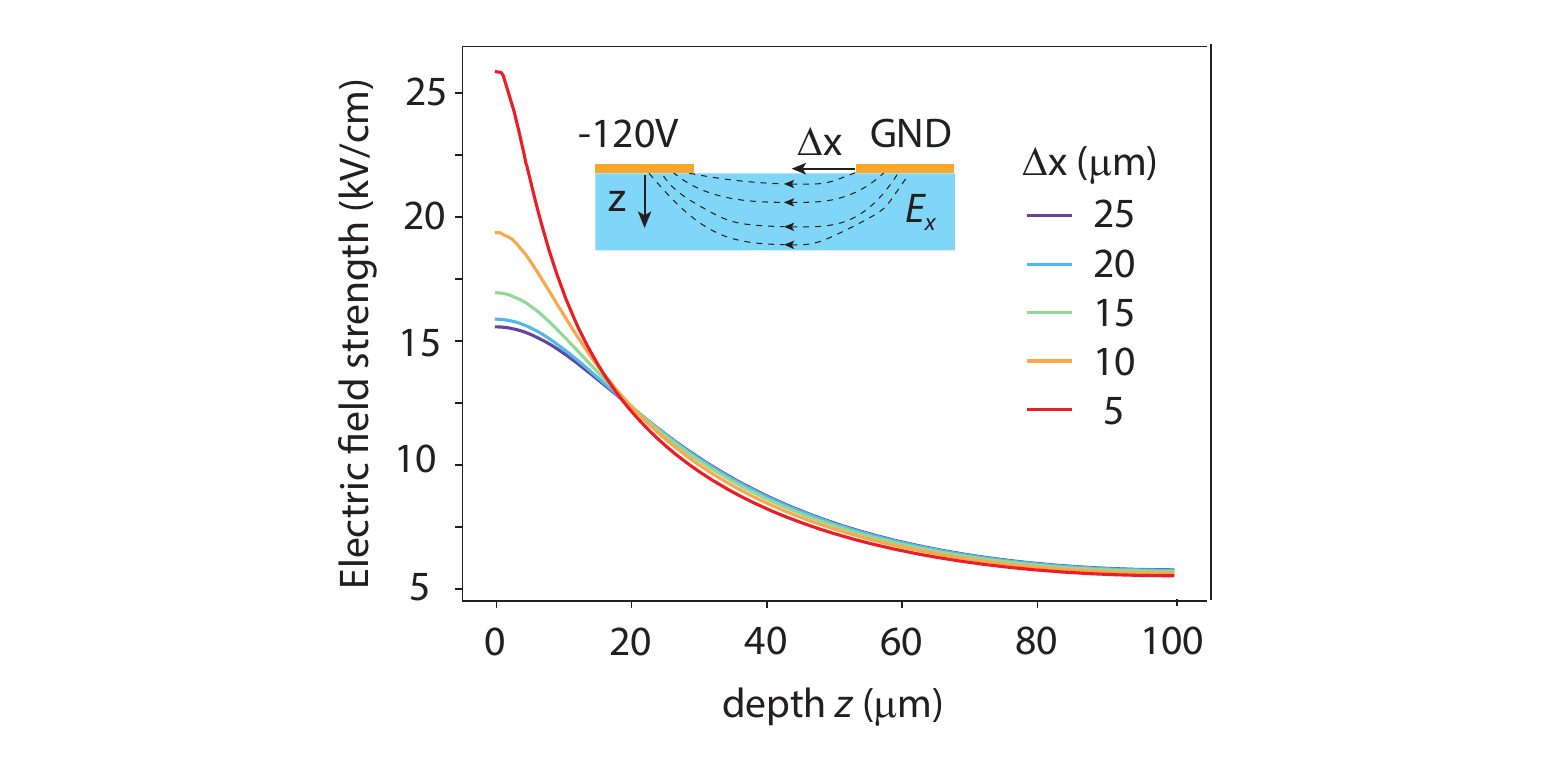}
	\caption{Simulated electric field strength $|E_x|$ as a function of depth into the diamond $z$ for various distances $|\Delta x$ from the ground electrode. All experiments conducted in this work were within 20\,$\upmu$m of the surface.}
	\label{fig:figs3}
\end{figure}

To estimate the electric field geometry, we used the COMSOL Multiphysics Electrostatics module. We recreated the geometry of the electrodes placed on the surface of the diamond and used the in-built diamond material which assumes the relative electric permittivity of diamond is $5.7\varepsilon_0$. One electrode was held at a potential of -120\,V and the other at ground. Figure \ref{fig:figs3} shows how the electric field strength ($\left| E \right|$) varies with depth inside the diamond. All points were calculated along a line passing through the centre of both electrodes. The 25\,$\upmu$m line is at the mid-point between the electrodes. The 5\,$\upmu$m point is closest to the grounded electrode. Only the electric fields on the side of the grounded electrodes were calculated as the electric field is symmetrical. It should be noted that the simple model described here does not include surface or internal, N-dominated Debye screening effects by charges diamond created via laser illumination, which reduces the electric field strength at the location of the NVs~\cite{oberg_solution_2020s, barson_nanoscale_2021s}.

\begin{figure}[t!]
	\centering
		\includegraphics[width = \textwidth]{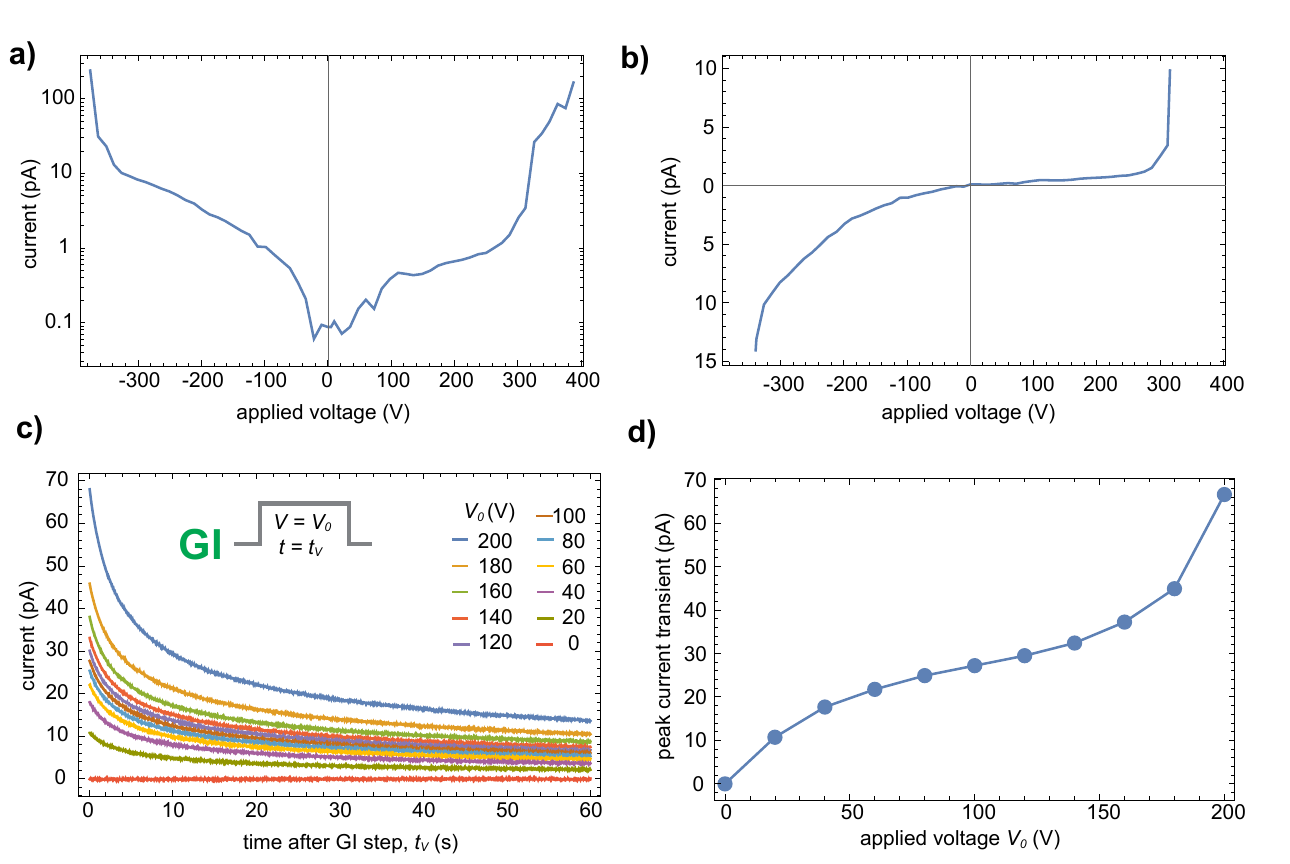}
	\caption{Voltage-vs-current characteristics for the diamond photoelectric device used in experiments. a) logarithmic scale, where the current magnitude at negative bias voltages is plotted, and b) detail on a linear scale of the low-voltage region, showing low leakage current. c) Current measured through device immediately following green initialisation scanning (GI) and application of a voltage $V_0$ for time $t_V$, and d) peak current transient vs applied voltage $V_0$. }
	\label{fig:figs4}
\end{figure}

The diamond sample equipped with metallic electrodes constitutes an electrical device that is composed of a metal-semiconductor-metal junction, each of which form Schottky-type barriers. The electrical characteristics of such a device exhibits a characteristic V-I curve at low bias voltages, where a very small leakage current flows: while one Schottky junction is forward-biased, the other is reverse biased and this limits the overall current. At higher bias voltages, the measured current abruptly increases, and for the voltages we can apply without damaging the device ($<$400\,V) we observe no saturation. A measured I-V curve of the device used in the experiments described in the main text is depicted in Fig. \ref{fig:figs4}. After a prolonged exposure to higher voltages and significant current flow, the I-V curves are observed to change, though gradually return to that depicted in Fig \ref{fig:figs4}. 

Though much information, including the barrier heights and ideality factors for each junction can be deduced from I-V measurements~\cite{grillo_currentvoltage_2021s}, it should be noted that the I-V curves represent the electrical characteristics of the \emph{entire} 500\,$\upmu$m of the diamond-metal contact rather than the comparatively smaller spatial region probed in  experiments (80$\times$80$\,\upmu$m). In principle, one could additionally study how the IV curves respond to the various optical initialisation protocols conducted in this work, though the presence of an unexplained systematic vexes these measurements. Following optical scanning (either GI or DP were observed to have the same effect) and application of a bias voltage, we observe a significant current transient with a slow decay time, typically of order 30\,s (Fig. \ref{fig:figs4}(c)). The magnitude of the current transient is observed to depend on the applied electric field, as shown in Fig. \ref{fig:figs4}(d), though the time constant is observed to be nearly completely independent, with time spent at zero electric field having the same effect as application of an electric field. We performed experiments to determine if this current transient had any role in the photocurrent measurements or spatial features in confocal measurements we study in the main text by pausing for 180\,s after optical preparation and application of the electric field, and found no significant difference in confocal images or photocurrent time dependence.    

\section{Optical preparation of nitrogen charge state}
As the most abundant defect in our diamond (1\,ppm), substitutional nitrogen is the prime candidate for the nonfluorescent defect that dictates the photoconductivity of the sample. All NV-hosting diamonds contain nitrogen to varying degrees, as it is an essential precursor in the formation on the NV centre itself and the negative charge state of the NV centre is hypothesised to arise from a nitrogen donor~\cite{loubser_electron_1978s, doherty_nitrogen-vacancy_2013s}. Three charge states of N$_\text{S}$ have been reported, though only N$^+$ and N$^0$ exist stably at room temperature in the bulk of the diamond~\cite{ashfold_nitrogen_2020s}, the negatively-charged state N$^-$ having only a 10\,ns lifetime~\cite{ulbricht_single_2011s}. Early photoconductivity measurements put the N$^0$ donor level at the commonly accepted value of 1.7\,eV below the conduction band~\cite{farrer_substitutional_1969s}, in agreement with ab initio modelling~\cite{bourgeois_enhanced_2017s}. However, more recent measurements reported low ionisation cross sections below 2.2\,eV, the now commonly reported `photoionisation threshold' for N$^0$~\cite{nesladek_dominant_1998s, rosa_photoionization_1999s, ashfold_nitrogen_2020s}. The difference is attributed to the additional energy required for rearrangement of the N and C atoms upon ionisation~\cite{bourgeois_enhanced_2017s}. Practically, this results in minimal charge interconversion of N under red (1.95\,eV) light.  

\subsection{Generation of N$^0$ and NV$^0$: green initialisation}
We consider first the green initialisation scan (GI), where a high power (8\,mW) green laser is scanned over the entire region of interest between the electrodes. The laser focus is approximately 7.5\,$\upmu$m below the surface of the diamond, and corresponds to an intensity of order 1\,MW/cm$^2$. For this high-intensity regime, charge cycling of the NV is very strong and results in rapid generation of free holes and electrons, which are captured by any NV$^-$ and N$^+$ surrounding the laser spot. Weaker processes, such as NV$^0$ electron capture and N$^0$ hole capture are not explicitly forbidden, though any generation of NV$^-$ or N$^+$ is quickly suppressed by the Coulombic attraction of holes and electrons, respectively. Under the laser spot itself, a steady state charge distribution is established with a significant NV$^-$ population, as evidenced by the bright spot at the centre of the CCP. The details of this charge distribution are unimportant, since scanning the laser beam (quasistatically on the $\sim$MHz timescales of ionisation and recombination) quickly swamps the bright areas with charge carriers generated under the new laser focus that subsequently result in NV$^0$ and N$^0$. While only NV$^-$ population can be definitively inferred from confocal measurements (at the noise floor after only one green scan), it is reasonable to assume a significant fraction of the nitrogen charge state population is converted to N$^0$ under the sustained generation of electrons from NV charge cycling. The 3D nature of the CCP generation means the NV$^0$/N$^0$ generation extends some 50\,$\upmu$m into the bulk of the diamond. The GI step is also effective at deleting old patterns imprinted on both the NV and N charge states, and thus forms the first initialisation step in every one of our experiments.

\subsection{Generation of N$^+$ and NV$^-$: defocused preparation}
We found that defocusing the green laser (while kept at the same power of 8\,mW) resulted in an increase in NV$^-$ generation. While not in itself evidence of a similarly dramatic change in the N$^0$ population, the two processes are intimately linked. For the NA = 0.6 objective lens used in our experiment, raising the objective lens 50\,$\upmu$m above the imaging plane (i.e. where the GI step was performed) reduces the optical intensity by a factor of 1900, but reduces the two-photon charge cycling rate of the NV ($\propto I^2$) by a factor of 3$\times 10^6$, effectively suppressing it compared to the single-photon ionisation rate of N$^0$. The reduced optical intensity, extended interaction time and inhomogeneous intensity profile of the light is the key to the DP step. An equivalent process to the formation of a CCP occurs, albeit with one-way photoionisation of N$^0$ being the dominant effect and this time, the steady-state charge distribution under the laser spot is the important part. 

Under sustained, low intensity green illumination, single-photon ionisation of N$^0$ occurs. The electrons generated by this process either 1) recombine with the host N$^+$ (geminate recombination), 2) diffuse some distance before capture by another N$^+$ or 3) be captured by a trap that is optically inactive. For cases 1 and 2, subsequent photoionisation always occurs, resulting in multiple capture and ionisation cycles between N$^+$ centres before eventual pumping into traps. Electron capture by NV$^0$ is weak due to the lack of coulombic attraction, but over the multiple seconds of exposure to light such processes may become significant. Once an NV$^0$ has captured the electron, the low optical intensity suppresses ionisation and extends the trap lifetime, resulting in efficient NV$^-$ generation. Further, without NV charge cycling occuring, hole generation and diffusion is assumed to be negligible, meaning trapping by NV$^-$ is a dead-end for the electron. After saturation of the NV and potentially other traps, the end result is local N$^+$ enrichment. As discussed in the main text, the exact charge composition of other defects is currently unknown. Defining a charge environment composed of only neutrally-charged defects, we consider a mesoscopic charge-neutral situation, with equal numbers of positively- and negatively-charged defects coexisting with no spatial structure. With this assumption, it is reasonable to assume that the generation of N$^+$ is accompanied by a concomitant increase in a negatively charged defect that is not the NV centre. This second inference stems from the observation that ionising the NV$^-$ population generated following a DP scan with red light (Main text, Fig 2(a)) has no bearing on the photoconductivity, and implies a neutral charge trap T$^0$ with a concentration significantly higher than the NV yet lower than substitutional nitrogen,
\begin{equation}
[\text{N}]>[\text{T}]\gg[\text{NV}].
\end{equation} 
Thus, the change in the N charge state after a DP scan is limited by the T$^0$ population rather than the NV population. The lack of an observed GR1 fluorescence feature from V$^0$ in this diamond (ZPL at 741\,nm \cite{grivickas_carrier_2020s}) raises doubts about this model. An alternative picture, mentioned in the Main Text, is that the electrons liberated from N$^0$ photoionisation in the DP step diffuse away from the illumination region, to be captured by other defects in the bulk or surface of the diamond, thus leaving a significant N$^+$-rich, charge-uncompensated region. Charge compensation thus occurs over a macroscopic region.

\subsection{Generation of N$^0$ and NV$^-$: Weak green preparation}
\begin{figure}[t!]
	\centering
		\includegraphics[width = \textwidth]{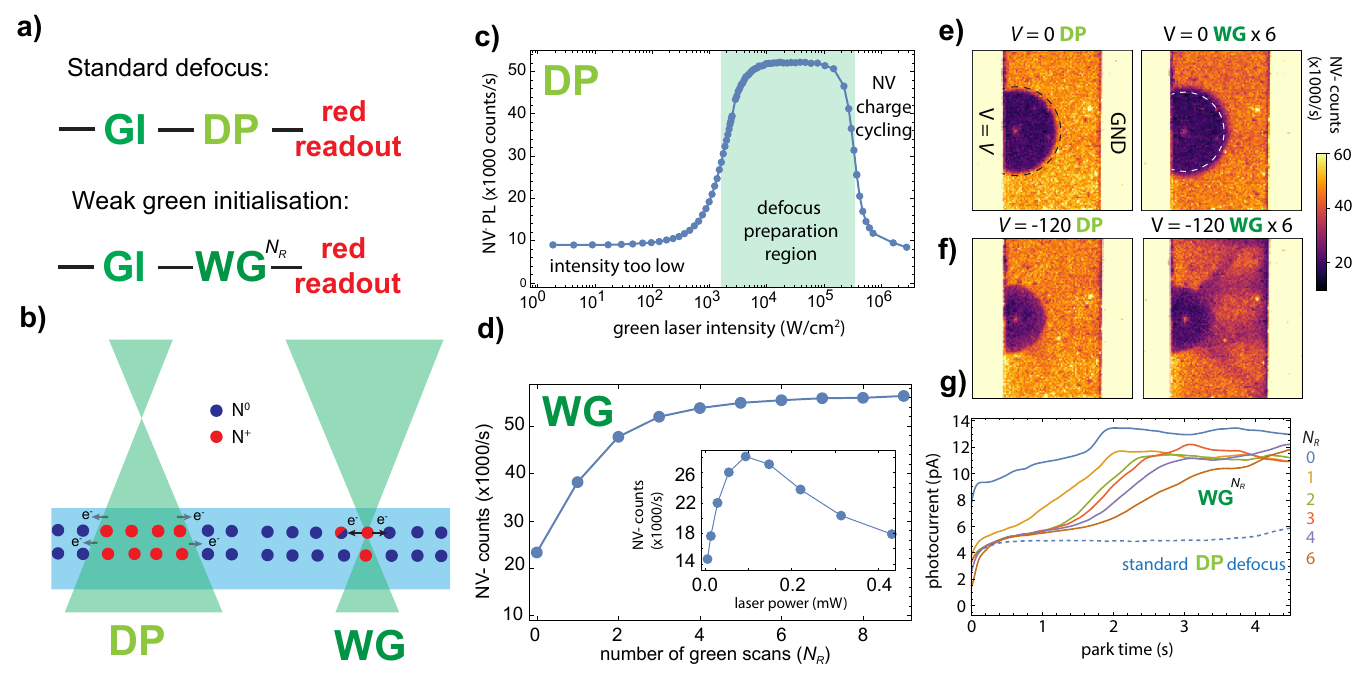}
	\caption{Defocused preparation vs. weak green scanning. a) Sequence of intialisation and readout steps. $N_R$ denotes the number of repeated green scans. b) Schematic depicting differences between the DP and WG sequences: while DP photoionises N$^0$ over an extended region, WG only does so under the laser spot. During a scan, the moving laser overwrites the previously prepared region with electrons, resulting in N$^0$ generation. c) Optimisation of defocused preparation. NV$^-$ PL counts are indicative of the NV$^-$ population generated in a fixed region of the diamond (a 30\,$\upmu$m circle between the electrodes). High intensities result in NV$^-$ charge cycling and hole generation, while low intensities do not photoionise N$^0$. In between, N$^0$ photoionisation is dominant over NV$^-$ charge cycling, resulting in N$^+$/NV$^-$ generation. d) Optimisation of WG: NV$^-$ generation vs number of repeated 90\,$\upmu$W scans, and (inset) power dependence for a single scan. We use $N_R = 6$ henceforth. e) Generation of a CCP with an 2\,s, 8\,mW laser pulse (no applied electric field) on a background prepared with DP (left) and 6 WG scans (right). All images are $(80\upmu\text{m})^2$ scans. The dashed lines depict the CCP radius in the opposing image: charges propagate slightly further under a WG preparation. f) Same as (e) but with $V = -120$\,V. While no filaments are present in the DP initialised background, clear filaments appear at earlier times with WG. g) Photocurrent vs time for an 8\,mW green laser park on standard DP background and with various $N_R$. }
	\label{fig:figs5}
\end{figure}

Reducing the laser power to a sufficiently low level and keeping the laser focus scanning in-plane should serve to emulate the defocused preparation step, assuming only the intensity of the light, rather than its spatial extent, matters. We compared weak green (WG) scanning to defocused preparation. With both initialisation steps preceded by a GI scan that is assumed to make N$^0$/NV$^0$ (Fig. \ref{fig:figs5})(a,b), we varied the intensity in the DP step by changing the height above the diamond and laser power to span 7 orders of magnitude in intensity and measured the PL within a 30\,$\upmu$m circle between the electrodes to assess NV$^-$ generation. Three regimes are apparent in Fig. \ref{fig:figs5}(c): when the intensity is too low, little to no photoionisation of N$^0$ occurs and thus no NV$^-$ population is generated. For high intensities, $I>10^5\,$W/cm$^2$, NV$^-$ charge cycling is strong and so hole generation and capture result in NV$^0$ formation, akin to a GI scan. In between, $10^3<I<10^5$\,W/cm$^2$, NV$^-$ generation is maximised due to photoionisation of N$^0$ and electron capture by NV$^0$. We typically used intensities of $10^4$\,W/cm$^2$ in DP scans.

Optimising the WG scan for NV$^-$ generation was achieved by first reducing the green laser power to find the point at which NV$^-$ generation was highest in a single scan ($90\,\upmu$W), and then repeating the scan six times (Fig. \ref{fig:figs5})(d). We then compared formation of a CCP (8\,mW, 2\,s park) without (Fig. \ref{fig:figs5}(e)) and then with (Fig. \ref{fig:figs5}(f)) an applied electric field. We observed that the radius of the CCP is slightly larger with WG preparation, which is assumed to generate more N$^0$ than N$^+$. Under a $-120$\,V voltage applied to the left-hand electrode, the difference between the two initialisations is far more apparent. While no filaments are discernable under DP preparation at 2\,s park time, with WG a clear V-shaped filament structure is resolved. In the main text we correlated formation of filaments with increases in photocurrent, and we clearly observe this here also. In Fig. \ref{fig:figs5}(g) we show the photocurrent vs time for DP initialisation (dashed blue line) and compare this for WG with various $N_R$. As the number of green scans increases, the photocurrent time dependence appears to smoothly change from the characteristic profile of purely GI initialisation (fast saturation, no plateaus) towards what we see with DP (plateaus and filaments). 

Changing the number of WG scans appears to be able to vary the N$^+$ concentration. This can be explained by invoking the model of nitrogen charge conversion in the macroscopic charge neutrality case, and would seem to confirm this model. Here, the optically-prepared region is limited to the micron-sized laser spot. N$^0$ is still ionised here, and NV$^-$ charge cycling suppressed, but with the assumption of macroscopic charge neutrality, electrons need only diffuse a a micron or so to escape the laser. As the laser scans, a similar overwriting effect that occurs with the GI scan occurs, except that this time only electrons are present, meaning a weak green scan prepares NV$^-$ (since no holes are generated in the wake of the scanning laser, NV$^-$ production is unaffected) and N$^0$ from electron capture by N$^+$. However, this inference is based on the assumed N$^0$ initial charge composition. Complete experimental verification of one model or the other is not readily apparent, and further work -- ideally using an independent metric for nitrogen concentration, such as electron paramagnetic resonance -- is needed to reveal the exact charge composition following DP scans.    
 
\section{Power, position and field strength dependence}

\subsection{Position of the photocurrent laser}
\begin{figure}[t!]
	\centering
		\includegraphics[width = \textwidth]{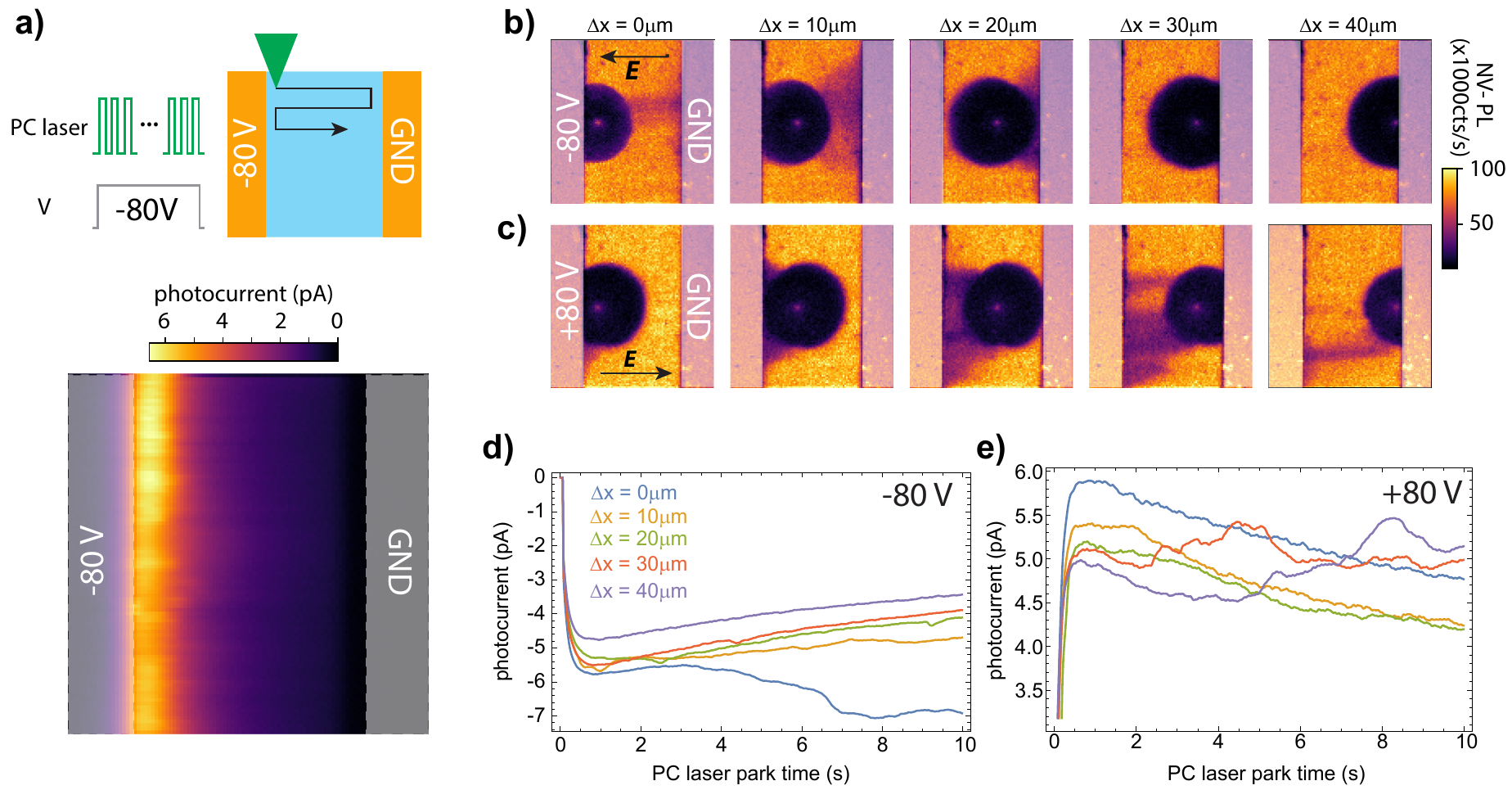}
	\caption{a) Scanning the laser beam while measuring photocurrent ($P$ = 8\,mW, $V = -80$\,V) reveals that the optimum point for maximised photocurrent generation is near the negative electrode.  b) Changing the laser park position from optimum ($\delta x = 0$) during a photocurrent measurement (after GI and DP initialisation) affects the spatial distribution of CCP and filament formation.  c) Same as (b) but with reversed electric field, here, the filaments form left to right. d) Time-dependence of the photocurrent for negative bias and (e) positive bias. All images are $(80\upmu\text{m})^2$ scans.}
	\label{fig:figs7}
\end{figure}

In all measurements reported in the main text, the photocurrent laser was positioned within 5-10\,$\upmu$m of the electrode held at $-120\,$V. This position was chosen as it corresponds to the point at which maximum photocurrent was detected when the laser position was scanned. Fig. \ref{fig:figs7}(a) depicts a scanning-photocurrent measurement of the inter-electrode gap, each line scan taking 4\,s to complete (20\,ms dwell time per 0.8\,$\upmu$m pixel plus dead time). We thus measure the steady-state photocurrent rather than the photocurrent at a given time as before, as no well-defined initialisation steps are taken before each scan. We measure maximum photocurrent yields near the negatively biased electrode, and such observations were also made in Ref. \cite{todenhagen_wavelength_2023s}. We now know from this current work that the increase in photocurrent near the negative electrode corresponds to hole current flowing in the `breakdown regime' from the ground electrode towards the CCP. This is observed vividly in the following experimental data. 

We studied the spatiotemporal dynamics of photocurrent using the methods we developed in this current work, changing the position where the photocurrent laser is applied between the electrodes. Fig. \ref{fig:figs7}(b-e) shows the relevant data. Starting with green initialisation and defocused preparation, the photocurrent laser (8\,mW) is parked for a variable time and the resulting CCP distribution is imaged while photocurrent simultaneously detected. When the photocurrent laser is parked close to the more negatively biased electrode, filaments form from the ground electrode: from right to left when negatively biased and from left to right when positively biased, in the direction of the applied electric field. As the laser park position is moved from the negative electrode, the CCP radius is observed to increase, and distort towards the negative electrode. These observations can be understood by noting that when the CCP overlaps with the negative electrode, photogenerated holes that would otherwise contribute to the CCP are drawn under or into the electrode. As the laser position is moved further away, holes are still drawn towards the negative electrode, resulting in a distorted shape, but now more holes contribute to the expanding front of the CCP. The time-dependence of the simultaneously measured photocurrent reflects the formation of filaments observed in images. When filaments are present, the photocurrent is observed to first rapidly increase, plateau, and then increase again upon the formation of filaments. When no filaments are visible we observe only the rapid initial increase and plateau. In both cases we also observe the magnitude of the photocurrent to slowly drift down, consistent with the time dependent transient reported in Sec. \ref{sec:dev_chars}. We note here that we have reverted to signed photocurrents to reflect the different device biases, whereas in the main text we plot the absolute value of photocurrent.

\begin{figure}[t!]
	\centering
		\includegraphics[width = \textwidth]{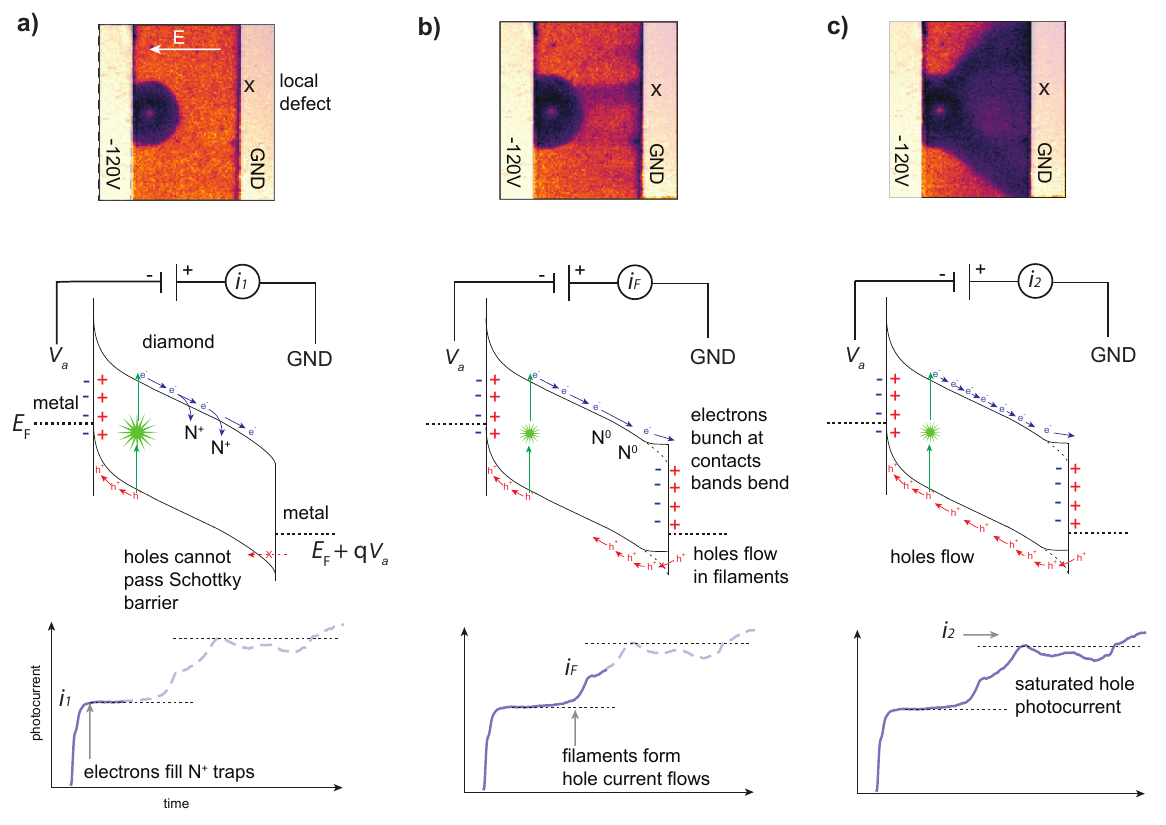}
	\caption{Photocurrent generation explained in terms of Schottky barriers. a) Application of the photocurrent laser near the negatively-biased electrode generates electrons and holes via NV charge cycling. Holes are swept towards the left and electrons to the right, and a photocurrent $i_1$ is established. As electrons pass through the diamond towards localised point defects on the right hand electrode, N$^+$ is converted to N$^0$ via electron capture. Holes cannot flow from the right to the left since the Schottky barrier is too high, both junctions are reverse-biased for charges on the metallic electrodes. b) Photoelectrons generated by the laser are trapped at the diamond-metal interface and form a space charge, which results in localised band bending and reduction of the Schottky barrier height for hole injection. Holes enter the diamond and flow along the channels of N$^0$ created by the electron current, which we observe as filaments of NV$^0$ and photocurrent begins to increase (to $i_F$). c) following a steep increase in photocurrent from hole injection, holes now flow through the diamond and a saturated electron-hole photocurrent $i_2$ is reached.}
	\label{fig:figs9}
\end{figure}

Varying the position of the laser offers considerable insight into the photoelectric mechanisms taking place. From Fig. \ref{fig:figs7}, it is evident that filaments only occur between the CCP and the more positively biased electrode. We summarise the inferred dynamics of photocurrent in Fig. \ref{fig:figs9}. Since the filaments are interpreted as hole current flowing \emph{towards} the CCP, this implies holes crossing the metal-diamond interface. However, without the photoelectric laser generating charges inside the diamond, no such current flow is observed to take place. We hypothesised in the main text that the hole current is able to flow when electrons from the photocurrent laser create corridors within the N$^+$ regions that are focused on defective sites along the electrode. We identified the flowing photocurrent prior to instigation of filaments as electron current. Under reverse bias, electrons are unable to overcome the metal-diamond Schottky barrier (evidenced by the very low DC current flowing at these voltages), but appear to traverse the diamond-metal junction (\emph{i.e.} from inside the diamond, effectively forward biased) without difficulty. 

\begin{figure}[t!]
	\centering
		\includegraphics[width = \textwidth]{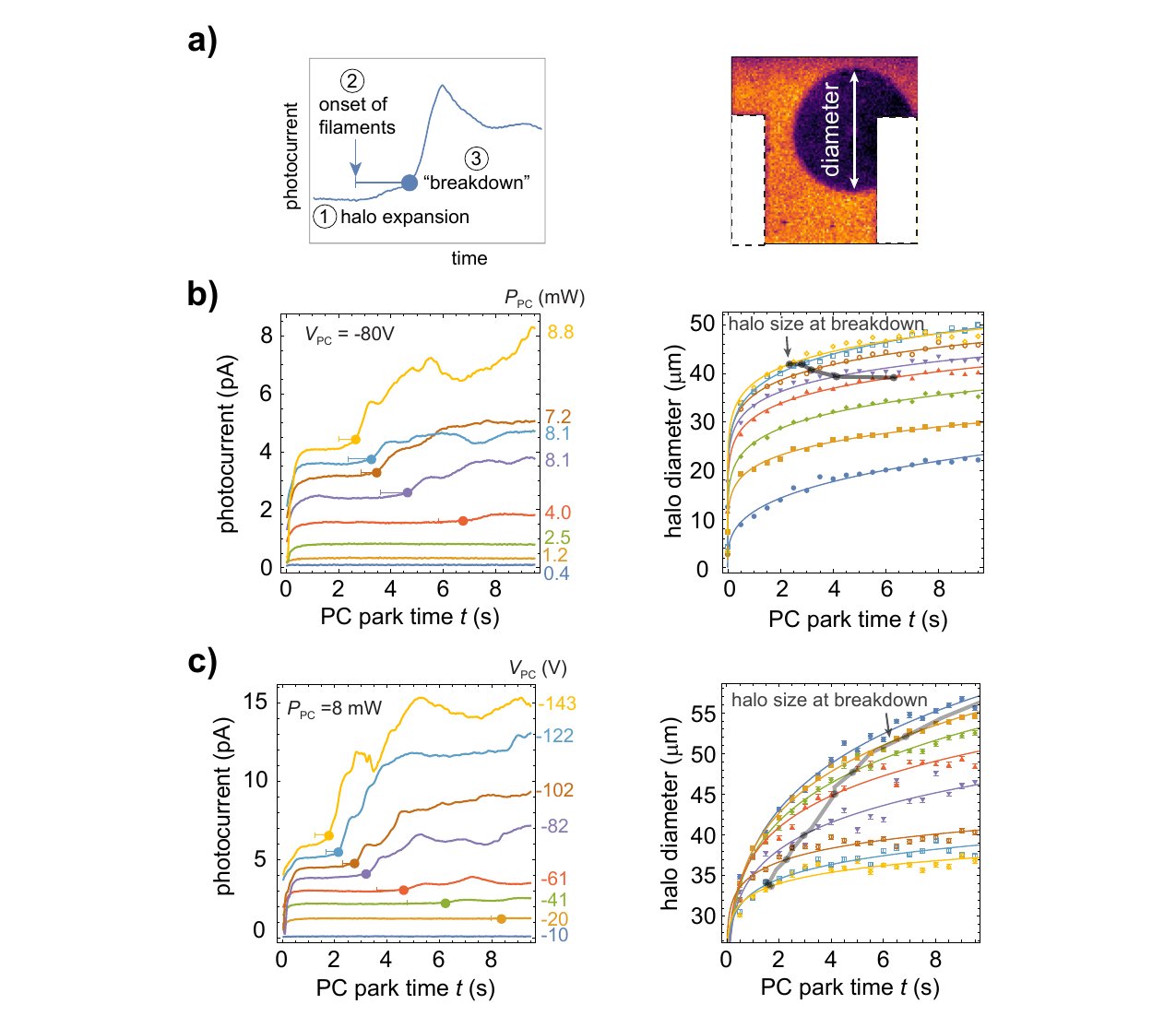}
	\caption{a) To study the effects on temporal and spatial photocurrent dynamics, we quantify the time corresponding to the onset of filaments (left) and measure the $y$-extent of the CCP (right). b) Increasing the laser power is observed to result in faster growth of larger CCP patterns (right) and consequently larger photocurrents and earlier onset of filaments and breakdown (left). At this applied electric field, (-80\,V), when the CCP diameter approaches $35-40\,\upmu$m, filamentation and subsequent breakdown occurs (dark trace). c) Increasing the electric field strength results in increased photocurrent and earlier onset of filamentation and breakdown. This is accompanied by a reduction in the CCP diameter, consistent with more charge carriers (holes) being swept towards the negative electrode. Overlaid on the right-hand panel is the CCP diameter when filamentation and breakdown occur. Solid lines on the right in (b), (c) are fits of the form $A e^{-\gamma t ^{0.25}}$.}
	\label{fig:figs8}
\end{figure} 
The model depicted in Fig. \ref{fig:figs9} can then be used to explain the data in Fig. \ref{fig:figs7}(a-c). When the photocurrent laser is parked near the more positively-biased electrode, no filaments or breakdown are observed, and the photocurrent even in the steady state is less. In this case, holes drift slightly towards the negative electrode, but always experience an N$^+$ rich environment: with no electrons to passivate the N$^+$, hole mobility is reduced and they do not flow across the device. Without a space charge growing near the negative electrode to slightly forward-bias the Schottky junction, no electron injection occurs either. One might expect that at sufficiently high voltages, electrons can overcome the Schottky barrier and flow into the diamond, \emph{i.e.} facilitating reverse filament formation, from photocurrent laser (generating holes) to the electrode. Electron injection at defective points along the electrode would be expected to nucleate filaments. However, an important distinction must be drawn here. The characteristic pattern of holes converging upon the photocurrent laser illumination point is a consequence of the paths through the N$^+$ space-charge region created by electrons that are generated at a point source. We never saw any filament-like features following current excursions at high bias voltages, i.e. in Fig. \ref{fig:figs4}.       

\subsection{Photocurrent laser power and electric field}
We then studied the spatiotemporal dynamics of photocurrent as a function of the two most pertinent parameters at our disposal, namely, laser power and electric field strength. We focus our analysis on the onset time of filament formation and then subsequent `breakdown' that are apparent from the photocurrent trace, and correspondingly the diameter of the CCP which is deduced from confocal images following the photocurrent measurement as shown in Fig. \ref{fig:figs8}(a). We define the CCP diameter as the FWHM extent of the dark charge pattern in the $y$-direction, since the strong distortion accompanying a breakdown event precludes reliable extraction of a characteristic horizontal spatial extent. While the halo is not necessarily radially symmetric (though approximately so in the space-charge dominated regime we are in), the $y$-diameter nevertheless constitutes a useful figure of merit to quantify the physical extent of the charge capture region. 

As the laser power is increased, we observe that the CCP grows faster, and subsequently that filaments and breakdown occur at earlier times, as shown in Fig. \ref{fig:figs8}(b). Indeed, when the CCP diameter nears 35-40$\,\upmu$m this corresponds to the onset of filament formation and subsequently breakdown. This observation implies that the formation of filament channels (at a given electric field strength) are mediated by the proximity of the CCP boundary to the ground electrode. When the electric field strength is increased in Fig. \ref{fig:figs8}(c), we again observe that breakdown events happen at earlier times, however, the charge capture halo diameter is observed to saturate to smaller values. We interpret this latter observation to essentially the same phenomenology discussed in Fig. \ref{fig:figs7}(b,c), with the stronger electric field pulling holes from the CCP into the electrode, thus preventing the expansion of the CCP.

\end{document}